# The ERICA Switch Algorithm for ABR Traffic Management in ATM Networks

Shivkumar Kalyanaraman[1], Raj Jain, Sonia Fahmy,Rohit Goyal and Bobby Vandalore

The Ohio State University, Department of CIS

Columbus, OH 43210-1277

Email: *shivkuma*@ecse.rpi.edu, {*jain, fahmy, goyal, vandalor*}@cis.ohio-state.edu

**Abstract**

We propose an explicit rate indication scheme for congestion avoidance in ATM networks. In this scheme, the network switches monitor their load on each link, determining a load factor, the available capacity, and the number of currently active virtual channels. This information is used to advise the sources about the rates at which they should transmit. The algorithm is designed to achieve efficiency, fairness, controlled queueing delays, and fast transient response. The algorithm is also robust to measurement errors caused due to variation in ABR demand and capacity. We present performance analysis of the scheme using both analytical arguments and simulation results. The scheme is being implemented by several ATM switch manufacturers.

## 1 Introduction

The ATM Forum Traffic Management Specification provides in precise details the rules for the source and destination end system behaviors for the available bit rate (ABR) service for asynchronous transfer mode (ATM) networks [1]. The switch behavior, however, is only coarsely specified. This provides the flexibility for various vendors to implement their own switch rate allocation algorithms. Several switch algorithms have been developed [2]-[9]. This paper describes one of the earliest switch algorithms.

The Explicit Rate Indication for Congestion Avoidance (ERICA) algorithm was presented at the ATM Forum in February 1995. Since then, its performance has been independently studied in many papers [4, 5, 7], and we have incorporated several modifications into the algorithm [9, 10]. This paper provides a consolidated description and a performance analysis of the scheme. The paper is organized as follows.

[1]Shivkumar Kalyanaraman is now in Dept. of ECSE, Rensselaer Polytechnic Institute, Troy NY 12180-3590

We begin by briefly examining the ABR service. In section 3, we describe basic concepts such as the switch model and design goals. Section 4 describes the algorithm. Section 5 presents representative simulations to show that the scheme works under stressful conditions; we also present analytical arguments in appendix A. Finally, our conclusions are presented in section 6.

## 2 The ABR Control Mechanism

ATM networks offer five classes of service: constant bit rate (CBR), real-time variable bit rate (rt-VBR), non-real time variable bit rate (nrt-VBR), available bit rate (ABR), and unspecified bit rate (UBR). Of these, ABR and UBR are designed for data traffic, which has a bursty unpredictable behavior.

The UBR service is simple in the sense that users negotiate only their peak cell rates (PCR) when setting up the connection. If many sources send traffic at the same time, the total traffic at a switch may exceed the output capacity causing delays, buffer overflows, and loss. The network tries to minimize the delay and loss using intelligent buffer allocation [15], cell drop [16] and scheduling, but makes no guarantees to the application.

The ABR service provides better service for data traffic by periodically advising sources about the rate at which they should be transmitting. The switches monitor their load, compute the available bandwidth and divide it fairly among active flows. This allows competing sources to get a fair share of the bandwidth and not be starved due to a small set of rogue sources. The feedback from the switches to the sources is sent in Resource Management (RM) cells which are sent periodically by the sources and turned around by the destinations (see figure 1).

The RM cells contain the source current cell rate (CCR), and several other fields that can be used by the switches to provide feedback to the source. These fields are: Explicit Rate (ER), Congestion Indication (CI) Flag, and No Increase (NI) Flag. The ER field indicates the rate that the network can support at the particular instant in time. When starting at the source, the ER field is usually set to the PCR, and the CI and NI flags are clear. On the path, each switch reduces the ER field to the maximum rate it can support and sets CI or NI if necessary [12].

The RM cells flowing from the source to the destination are called forward RM cells (FRMs) while those returning from the destination to the source are called backward RM cells (BRMs). When a source receives a BRM, it computes its allowed cell rate (ACR) using its current ACR, CI and NI flags, and the ER field of the RM cell [13].

# 3 Basic Concepts

In this section, we introduce the switch model used in the rest of the paper, as well as a description of key goals for the switch algorithm design.

## 3.1 Switch Model

Our switch model is shown in figure 2. Every service class has a separate FIFO output queue which feeds to the output link under the control of a scheduling mechanism. The ERICA algorithm works at the ABR output queuing point. We do not discuss aspects of the scheduling mechanism, except for the fact that it provides the switch algorithm with the knowledge of the available capacity for the ABR service. We assume that there are at most two classes (VBR and ABR) and ABR has the lowest priority, i.e., it gets the leftover capacity after VBR cells are transmitted. In practice, it is desirable to allow some minimum capacity for processing aggregate ABR traffic when there is contention.

In this paper we do not deal with the case of individual ABR VCs having guaranteed non-zero Minimum Cell Rates (MCRs) [1]. Techniques for adapting a scheme for MCR are suggested in [17]. Other issues not addressed in this paper include the effect of complex queueing strategies like per-VC queueing, network segmentation using the Virtual Source/Virtual Destination (VS/VD) option [1], and point-to-multipoint ABR connections. Some of these issues are addressed elsewhere [9, 19].

In figure 2, observe that the RM cells of a connection enter the switch through one port in the forward direction (with the forward data flow) and exit through another port in the reverse direction (with the reverse data flow). In the ERICA algorithm, we monitor the forward flow for metrics, but give the feedback in the backward RM cells, thus minimizing the latency in delivering feedback to sources. We measure certain characteristics of the flow over intervals of time, called *"switch measurement intervals"* **or** *"switch averaging intervals."* The measured quantities are placed in a table for use in the reverse direction when calculating feedback. The feedback calculation may be performed when a backward RM cell (BRM) is received in the reverse direction, or may be pre-calculated at the end of the previous averaging interval (of the forward direction port) for all active sources. The latter option may also be implemented using lazy evaluation and/or in the background using a dedicated processor.

One key feature of the ERICA scheme is that it gives *at most one feedback value per-source during any averaging interval.* As a result, it precludes the switch from giving multiple conflicting feedback indications in a single averaging interval using the same control values. Further, since there can be multiple switches on a VC's path, the allocation given to the source is the minimum of all the switch allocations. For performance purposes, it is desirable to have all the switches implement the same switch algorithm [20], but the traffic management standard [1] does allow switches from multiple vendors to interoperate.

While ERICA gives feedback in the explicit rate field in the RM cell, it is possible to additionally throttle or moderate the sources by setting the CI and NI bits in the RM cell using policies suggested by several other schemes [4, 17]. Also, in our studies we set the Rate Increase Factor (RIF) parameter to one allowing maximum increase. Sources/switches can choose to be more conservative and set it to lower values.

## 3.2 Design Goals

In this section, we enumerate the design goals of ERICA in a *priority order.* The goals describe the desired "steady state" operation and the priorities are used for a graceful degradation of the goals under transient conditions. Specifically, under such conditions, lower priority goals are traded off to achieve higher priority goals. Besides this list of goals, the algorithm also aims at stability and robustness. A mathematical formulation of the rate-based control problem may be found in [21] and references therein.

1. **Utilization:** Maximize link utilization $\rho(t)$. Since the capacity of a link is potentially shared by several classes, and the switch algorithm controls only the ABR class, we translate this goal into one of maximizing the ABR utilization.

The ERICA scheme tracks ABR utilization using a metric called "load factor" ($z$, see section 4.1). Roughly, $z$ is the ratio of the ABR input rate and the ABR capacity. The utilization goal of ERICA is then a *steady state* operating point in the neighborhood of $z = 1$. If the system has *no steady state*, ERICA gives highest priority for maximizing the average ABR utilization. For the latter case, note that if the aggregate demand is consistently small, maximum utilization cannot be achieved. Further, the utilization metric gives only a partial picture of efficiency. Ideally, an "efficient" scheme should also control queuing delays within acceptable limits (discussed next).

2. **Queuing Delay:** As mentioned above, link or ABR utilization is only a partial indicator of "efficiency." Specifically, $\rho(t)$ or $z$ could be unity, while a huge queue backlog exists at the bottleneck. An important goal is to achieve a target queuing delay at the bottleneck. The combination of the delay and utilization goal is often called **"congestion avoidance."**

   Figure 3 shows the *target* throughput-delay behavior of our system. Throughput is used in the figure because it reflects the performance seen by the application layer. We assume an infinite buffer and do not model the effects of cell losses. The figure shows that for small loads, throughput of the system should increase linearly, while delay is almost constant. The throughput saturates at a point called the *"knee,"* while delay increases linearly. When the delay is too large, the throughput is expected to degrade due to timeouts at higher layers and wasteful retransmissions. This point is called the *"cliff,"* and is an unstable point we never want the system to reach. Specifically, we do not want the system to have unbounded queuing delays. Further, the peak queuing delay due to transient load changes should be minimized, and the queues built up drained quickly to reach a target steady state value.

   The target steady state operation of the system is at a point between the knee and the cliff. In this state, the system has a "pocket of queues" which can be useful in keeping a link temporarily utilized when capacity suddenly becomes available. Note that when the system is in a state of high variation, we may never achieve the target queuing delay because the system itself has no steady state. Under these conditions, the aim is to keep the average utilization high, the average queue size small, and prevent queuing delays from becoming unbounded.

3. **Fairness:** The next priority of ERICA is to allocate rates in a fair manner. One commonly used criterion for describing fairness is the *max-min allocation* [6]. Intuitively, it calls for maximizing the allocation of the minimum rate source, i.e., to give each contending source a "maximum possible equal share" of the bandwidth. In a configuration with $n$ contending sources, where the $i$th source is allocated a bandwidth $x_i$, the allocation vector $\{x_1, x_2, \ldots, x_n\}$ is said to be feasible if all link load levels are less than or equal to 100%. If we order the elements of the vector in ascending order, then the max-min allocation is defined as the vector which is lexicographically the largest among the set of vectors. For the proof given in appendix A, we use the notion of the maximum possible equal share, with the feasibility constraint that the load factor ($z$, see section 4.1) is in the neighborhood of unity. The ATM Forum has also specified several fairness definitions for cases when ABR VCs have non-zero Minimum Cell Rates (MCRs) [1].

   We believe that fairness is a long-term concept, i.e., if two sources send traffic for a sufficiently long time (like two large file downloads), then their average rates of transfer should be max-min. During system transients, the max-min balance may be disturbed, but the system should converge back to max-min rates if no further transient conditions occur. However, a service class such as ABR (whose capacity and demand vary dynamically) may *always* be in a transient state. If the system has only persistent sources, but its capacity varies, fairness can be measured over the long-term using average source rates. However, if the demand fluctuates as well, there is currently no known model for quantifying fairness.

The switch algorithm is also designed to meet the following goals (which do not assume any priority order):

**Stability and Transient Performance:** Given that the scheme can reach steady state operation, it should be robust to variations which cause deviations from the steady state. A system which can re-establish its steady state after perturbations is said to be *stable.* The *transient performance* of the scheme determines how quickly steady state is re-established.

**Robustness:** In the case that the system has *no steady state* (e.g., due to persistant variation in capacity and demand), the scheme should be *robust*, i.e., its essential performance metrics should not degrade to unacceptable levels. For example, the queue length should not be unbounded, or the average utilization should not be very low if the average demand is not low. In ERICA, the priority of the goals stated above determines which goal is traded off under such stressful conditions.

This paper does not address the issues of scheduling, policing, buffer allocation, cell drop policies and interaction with transport layers in this paper. References [9, 15] look at some of these issues in greater detail.

# 4 The ERICA Algorithm

The ERICA algorithm operates at each output port (or link) of a switch. The switch periodically monitors the load on each link and determines a load factor ($z$), the ABR capacity, and the number of currently active virtual connections or VCs (N). A measurement or "averaging" interval is used for this purpose. These quantities are used to calculate the feedback which is indicated in RM cells. Recall from the discussion in section 3.1 that the measurements are made in the forward direction, whereas the feedback is given in the reverse direction. Further, the switch gives at most one new feedback per source in any averaging interval. The key steps in ERICA are as follows:

**Initialization:**

MaxAllocPrevious ← MaxAllocCurrent ← FairShare

**End of Averaging Interval:**

$$\text{Total ABR Capacity} \leftarrow \text{Link Capacity} - \text{VBR Capacity} \quad (1)$$

$$\text{Target ABR Capacity} \leftarrow \mathit{Fraction} \times \text{Total ABR Capacity} \quad (2)$$

$$z \leftarrow \frac{\text{ABR Input Rate}}{\text{Target ABR Capacity}} \quad (3)$$

$$\text{FairShare} \leftarrow \frac{\text{Target ABR Capacity}}{\text{Number of Active VCs}} \quad (4)$$

$$\text{MaxAllocPrevious} \leftarrow \text{MaxAllocCurrent} \quad (5)$$

$$\text{MaxAllocCurrent} \leftarrow \text{FairShare} \quad (6)$$

**When an FRM is received:**

CCR[VC] ← CCR_in_RM_Cell

**When a BRM is received:**

$$\text{VCShare} \leftarrow \frac{CCR[VC]}{z} \tag{7}$$

IF $(z > 1 + \delta)$

$$\text{THEN ER} \leftarrow \text{Max (FairShare, VCShare)} \tag{8}$$

$$\text{ELSE ER} \leftarrow \text{Max (MaxAllocPrevious, VCShare)} \tag{9}$$

$$\text{MaxAllocCurrent} \leftarrow \text{Max (MaxAllocCurrent,ER)} \tag{10}$$

IF (ER > FairShare AND CCR[VC] < FairShare)

$$\text{THEN ER} \leftarrow \text{FairShare} \tag{11}$$

$$\text{ER_in_RM_Cell} \leftarrow \text{Min (ER_in_RM_Cell, ER, Target ABR Capacity)} \tag{12}$$

The following sections explain the design of the above algorithm and its parameters, and also address issues of measurement of quantities such as the ABR capacity and input rate. A complete pseudo code of the scheme including all the features described below is provided in reference [9].

## 4.1 Load Factor

The load factor is the ratio of the measured input rate at the port to the target ABR capacity. According to equation (3):

$$z \leftarrow \frac{\text{ABR Input Rate}}{\text{Target ABR Capacity}}$$

In the above formula, the target ABR capacity is a fraction of the total ABR capacity (equation (2)). We assume total ABR capacity to be the link bandwidth minus the VBR bandwidth usage (equation (1)). In the steady state, ERICA aims at keeping the link utilization equal to the ratio of target ABR capacity and total ABR capacity. Under transient conditions, ERICA allocates the difference between total ABR capacity and target ABR capacity for draining the queues built up.

The load factor $z$ (also referred to as "overload factor" or "overload") is an indicator of the congestion level of the link. High values are undesirable because they indicate congestion; so are low values which indicate link underutilization. In the steady state, the bottleneck operates in the neighborhood of a load factor of unity ($z = 1$).

## 4.2 Achieving Max-Min Fairness

The ERICA algorithm calculates two quantities which are used to achieve max-min fairness. The first quantity is called "VCShare," which is calculated either at the end of an averaging interval and stored in a table, or calculated when a backward RM cell is processed to insert feedback. Equation (7) computes VCShare as:

$$\frac{CCR[VC]}{z}$$

CCR is the estimate of the source current cell rate. CCR may be read from the forward RM cells of the VC, or measured independently by the switch. Either way, the CCR value is stored in a table and used for this calculation. As shown in appendix A, this term helps the system converge to the neighborhood of $z = 1$ (utilization goal) within O(log N) cycles from any initial state. Note that a symmetric system (feedback delays identical), and unconstrained sources converges to such a state (neighborhood of $z = 1$) in a single cycle.

However, allocating sources their "VCShare" alone does not achieve max-min fairness. For example, when $z$ equals unity, assigning VCShare will not change the system state (even if it is unfair). A mechanism is required to equalize the rate allocations, while ensuring that the bottleneck load factor returns to the neighborhood of unity.

One possible mechanism is to calculate the *maximum* of the VCShare values and assign this maximum value to all sources. When the rates (CCRs) are *equalized* in this manner, subsequent VCShare values will be identical and the load factor will quickly converge to the neighborhood of unity within O(log N) cycles (see appendix A). One disadvantage of this mechanism is that it causes sharp load changes and extended periods of overload, which would result in undesirably large transient queuing delays. Hence, the ERICA algorithm uses a moderated version of this equalizing mechanism. We introduce a second term called "Fairshare," which guarantees a "minimum fairness" to contending sources. Fairshare is computed at the end of an averaging interval as in eqution (4):

$$\text{FairShare} \leftarrow \frac{\text{Target ABR Capacity}}{\text{Number of Active VCs}}$$

Intuitively, Fairshare is the *minimum share* which every active source deserves. ERICA allows any source sending at a rate below the $FairShare$ to rise to $FairShare$ at every feedback opportunity. Observe that if all sources converged to Fairshare, then the system is fair, and has a load factor of unity. ERICA moderates the load increase by not allowing a source currently sending below Fairshare to send more than Fairshare in the next cycle (equation (11)). This policy restricts the change in load (especially from small values) and allows the bottleneck one more cycle to stabilize its load. These features and mechanisms (VCShare, rate equalization, at least Fairshare, at most Fairshare if CCR low) are incorporated as steps in the ERICA algorithm as presented in equations (7) through (11).

The variables $MaxAllocCurrent$ and $MaxAllocPrevious$ are used to find the maximum allocation in one interval and to use this value in calculating the allocation in the subsequent interval (equations (5), (6), (9), (10)), . The parameter $\delta$ is used for the equalization of allocations (equation (9)) when the load factor is in the neighborhood of unity. It is set to a small value (0.05 to 0.1), because large values would result in extended periods of overload. The allocation written in the RM cell is the minimum of the calculated ER value, the value already in the cell (possibly written by other switches), and the target ABR capacity (equation (12)).

As we mentioned earlier, ERICA gives at most one new feedback per averaging interval. So the above feedback calculations could be done at the end of the averaging interval (possibly in the background) to speed up the RM cell processing. Also recall from section 3.1 that the CCR value used is read from the forward RM cells (or measured) and stored in the table for the calculations, as are the results of the computations. The feedback is inserted in the backward RM cells.

The description given so far is the core of the ERICA algorithm. Appendix A outlines a proof of convergence to max-min allocation. The proof uses the core algorithm, a single bottleneck, infinite sources, different round-trip times (RTTs), and switch averaging intervals at least as large as the maximum RTT of any VC through it (i.e. reliable measurement).

## 4.3 Queue Control

In section 4.1 we noted that the Target ABR Capacity is a fraction of the Total ABR Capacity. More generally, this fraction is a function of the queuing delay, f(Q), i.e.,
Target ABR Capacity = f(Q) $\times$ Total ABR Capacity.

The function f(Q), called the *"queue control function"* allows only a selected fraction of the available capacity to be allocated to the sources. The remaining capacity is used to drain the current queue. The original ERICA philosophy was that correct rate assignments depend more upon the aggregate input rate, rather than the queue length. However, three facts about queuing delays make them important to consider in feedback calculation: **a)** non-zero steady state queues imply 100% utilization, **b)** a system with very long queues is not operating efficiently, and, **c)** a service providing controlled queuing delay may help scalably support applications with delay requirements (e.g., variable quality video).

A simple queue control function is the constant function, i.e., a fixed parameter. We had used this function (called *"Target Utilization (U)"*) for the OSU scheme [8], and earlier versions of ERICA. We had found this function adequate in representative LAN (small round trip), and low error/variance WAN (large round trip) cases. The drawbacks of the constant function are as follows: it restricts the system utilization to a maximum of $U$ in the steady state; the system cannot achieve a queuing delay target; and it does not provide compensation when measurement and feedback are affected by errors.

The alternative is to have f(Q) vary depending upon the queuing delay. A number of such functions can be designed. One such function is the following (also see figure 4):

$$\text{f(Q)} = \frac{a \times Q0}{(a-1) \times Q + Q0} \quad for\ Q > Q0$$

and

$$\text{f(Q)} = \frac{b \times Q0}{(b-1) \times Q + Q0} \quad for\ 0 \leq Q \leq Q0$$

f(Q) is a number between 1 and 0 in the range $Q0$ to infinity, and between $b$ and 1 in the range 0 to $Q0$. Both curves intersect at $Q0$, where the value is 1. Observe that these are rectangular hyperbolic functions which assume a value 1 at $Q0$. This function is lower bounded by the queue drain limit factor ($QDLF$):

$$f(Q) = Max(QDLF, \frac{a \times Q0}{(a-1) \times q + Q0}) \quad for\ q > Q0$$

To emphasize the fact that we aim to control the queuing delay and to maintain parameter uniformity for different link speeds (and varying ABR capacities), we use a parameter $T0$ (target queuing delay) which is converted into the target queue length $Q0$ parameter before performing the calculation given above. Thus, the function f(Q) has four parameters: *T0, QDLF, a* and *b*. We chose this function because it has fewer points of discontinuity, smaller number of parameters required and is not too complicated to calculate. Threshold-based or hysteresis-based linear or constant functions are also possible. However, they require a large set of parameters which need to be re-tuned for different link speeds and distances.

The properties of our function f(Q) are as follows. It assumes a value of 1 at the desired steady state (utilization = 100%, queuing delay = $T0$). For delays smaller than $T0$, the hyperbola controlled by parameter $b$ (called $b$-hyperbola) is used to allow queues to accumulate till the target value. If the "pocket of queues" goal is not desired (e.g., in environments with high load/capacity variance), the $b$-hyperbola is not necessary, leaving us with just three parameters. This is the setting ($b = 1$) used in our simulation results. The parameter $T0$ specifies the target queuing delay, and also affects how quickly excess queues are drained. A larger $T0$ results in slower allocation of drain capacity. Hence $T0$ can be set to small values if the primary goal is to quickly drain excess queues.

The $a$-hyperbola, in conjuction with $T0$, determines how much "drain capacity" is used for draining out the queues built up. More drain capacity is allocated when the queue lengths are larger, up to a maximum of $(1 - QDLF)$. An important use of the queue control function is to compensate for measurement and feedback errors caused by system (load/capacity/source activity) variation. The parameter $QDLF$ defines the tolerance limit of the system to such variation. If variation is large, other techniques like the use of larger averaging intervals, and long-term averaging of metrics (see section 4.4) must be combined with queue control for ensuring robustness and effective control of the system. The queue control parameter choices are further described in section 5.1. Another alternative in this case (which we don't explore further) is to set the CI and NI bits if the queue lengths grow unacceptably high.

## 4.4 Robustness Issues

The performance of the ERICA algorithm depends significantly upon the way measurements are done and its parameters chosen. In this section, we discuss measurement aspects and describe scheme features that reduce errors in measurement. These features have been designed with scalability and high variance conditions in mind. Some of these features (e.g., averaging of metrics) need not be implemented in LANs or in WAN switch ports which do not support a large number of sources and/or do not exhibit a high variation in ABR capacity.

### 4.4.1 Switch Averaging Intervals

The ERICA algorithm measures certain quantities during consecutive averaging intervals. The question which immediately arises is how long each interval should be. Ideally, the interval length should be at least the maximum of the following: the longest round trip time and the time required to see at least one RM cell of every active source. This is the assumption used in the proof described in appendix A. While this assumption is reasonable for LANs where the round trips are short, it requires long intervals in WANs and satellite networks. Since ERICA gives only one feedback per averaging interval, links could experience long periods of underutilization due to lack of new feedback. The performance can be enhanced using shorter averaging intervals which could temporarily reallocate available capacity to nearby sources keeping the link utilization high, and then converge to max-min fairness when the long RTT source(s) respond. The drawback for shorter intervals is the possibility of variation (errors) in measurement.

Intervals can be of fixed length or of variable length (triggered when a fixed number of cells are received, M). The variable length intervals require special care to be taken when quantities like load factor or capacity (ratios) are averaged over multiple intervals [22]. As a rule of thumb, the interval length should be set such that it is possible to see at least one cell (preferably one RM cell) of every source during the interval, assuming that the minimum rate of any source does not fall below some low threshold. In addition, the interval must be long enough to smooth out high frequency oscillations in load and capacity. For example, the switch interval for a WAN/satellite switch port at OC-3 rates (RTTs in the range of tens/hundreds of milliseconds) can be fixed at about 5 milliseconds (which filters high frequency oscillations, and can allow identification of about 1500 sources each sending at a rate of 100 kbps).

### 4.4.2 Reliable Counting of Number of Active VCs

The ERICA algorithm requires that the number of active VCs be measured accurately. This is because it allocates every active VC *at least* Fairshare, i.e., target ABR capacity/N. Underestimation of N results in overallocation of rates. The ERICA algorithm can partially compensate for such errors through the queue control mechanism (which reduces the target ABR capacity). But, the queue control mechanism is limited by its parameters (notably QDLF) and hence cannot compensate for gross underestimation of rates. We describe a couple of techniques to reduce the error in estimating N.

The first technique, called "bidirectional counting," counts a VC as active if either a cell of the VC is seen in the forward direction, or a BRM is seen in the reverse direction. This technique is useful when traffic is bursty. Specifically, a VC can now be counted as active even when a BRM cell is seen in the reverse direction and no cells are seen in the forward direction (e.g., TCP traffic during startup).

Another technique to overcome the problem of underestimating the number of active VCs is to use exponential averaging to *decay* the activity level of a VC by a *DecayFactor* in each successive interval a VC is not seen. If a VC is seen during an interval, its activity level is reset to one (and not decayed). The *DecayFactor* parameter (ranges from 0 to 1) dictates a tradeoff: how soon is an idle VC considered inactive versus how much is N underestimated. Since the latter concern dominates when small intervals are used, a large fraction such as 0.9 is a reasonable choice in this case. Alternatively, the averaging interval could be set to be sufficiently long (trading off response time) and these techniques avoided altogether.

### 4.4.3 Load/Capacity Averaging and Boundary Cases

When the averaging interval is short, there may be cases when no input cells are seen during an interval, or ABR capacity changes suddenly. This results in a large variation in quantities used by ERICA (capacity and load factor), and consequently to highly varying and/or erroneous feedback. For robustness, these quantities should be averaged. An exponential averaging method is used for this

purpose:

Averaged ABR Capacity = $\alpha \times$ Measured ABR Capacity + $(1 - \alpha) \times$ Averaged ABR Capacity.

The load (input cells/sec) is also averaged this way, using the same parameter, $\alpha$. The choice of the parameter $\alpha$ is described in section 5.1. The load factor is then calculated as the ratio of the averaged load and the target ABR capacity (derived from the averaged ABR capacity). The exponential averaging technique can be implemented as described above given that the measurements (load and capacity) are done over *fixed* length averaging intervals.

For *variable* length averaging intervals, a minor variant of this technique is required. The reason is that ratios where the numerator and denominator both vary (e.g., load = number of cells input/ measurement interval length) cannot be averaged using simple arithmetic means or exponential averaging [22, 9].

Observe that the averaged load factor thus calculated is never zero or infinity unless the input rate or ABR capacity are always zero. The technique can be extended to optionally average other "rate" quantities measured at the switch. An example is the measurement and averaging of per-VC CCR at the switch (instead of reading from the the RM cell).

Finally, we have boundary conditions to cover certain extreme cases. First, the estimated number of active VCs should never be less than one (set to one otherwise). Second, when capacity is measured as zero, the feedback calculated is zero. Third, when input load is zero, the allocation is equal to the Fairshare.

# 5 Performance Evaluation of ERICA

In this section, we present simulations to verify the performance of ERICA under stressful conditions not considered by the analytical method used in the proof (see appendix A). All our simulations use the core ERICA algorithm (see section 4.2), the queue control mechanism and the measurement and averaging techniques described above.

## 5.1 Parameter Choices

The ERICA algorithm was designed to minimize the number of parameters to be chosen by the network administrator. The key parameters are the max-min fairness parameter ($\delta$), the length of the switch averaging interval, the queue control parameters ($T0$, $a$, $b$, $QDLF$) and the metric averaging parameters (if any).

The purpose and effect of the *max-min fairness parameter*, $\delta$, is described in section 4.2 and in appendix A. Briefly, $\delta$ determines the range of load factor values when the rate allocations are equalized to achieve max-min fairness. The algorithm reaches a steady state where its load factor remains in the range $[1, 1+\delta]$ (see appendix A). If $\delta$ is large, the queue control algorithm is strained to keep the steady state queues under control. If $\delta$ is too small, then the algorithm takes longer to converge to max-min fairness. We therefore recommend a range [0.05, 0.1] for $\delta$. Our simulations use the value 0.1.

The choice of the *switch averaging interval* involves a tradeoff between accuracy and reliability of measurement (longer intervals) versus the speed of response (shorter intervals). In addition, the choice of fixed length switch intervals (over variable length intervals) simplifies long term averaging for measured quantities. For switches in LAN environments (low ABR variation, small RTT and number of VCs), we recommend intervals of 1 to 5 ms (no long-term averaging required), and for WAN/backbone environments (high ABR variation, large RTT and number of VCs), fixed-intervals of 5 to 20 ms, with long-term averaging of measurements (N, capacity and load factor). Our simulations use the value $5ms$.

The following discussion of *queue control parameters* complements the discussion in section 4.3. We classify switch ports into LAN ports and WAN ports based on the length of the link they are connected to. In general, LAN switch ports could avoid implementing a complex queue control function and use a simple target utilization parameter (constant function) set in the range of, say, 0.8 to 0.95. WAN switch ports could implement the suggested queue control function (see section 4.3) since it is likely that ABR load/capacity variation will be high.

The parameter $b$ is chosen for LAN links to be in the range $[1, 1.05]$. This setting aims for a steady state target queuing delay. For WAN links, robustness assumes priority over steady state queuing delays and we recommend setting the parameter $b$ to one. This setting avoids implementing the $b$-hyperbola as well as avoids oscillatory behavior due to fluctuation of queues in the steady state. For the parameters $a$ and $QDLF$, we have found the parameter settings $a = 1.15$ and $QDLF = 0.5$ to be successful for a wide range of configurations [9]. Smaller values of $a$ and larger values of $QDLF$ reduce the capability of the scheme to respond to sudden queue increases. On the other hand, larger values of $a$ and smaller values of $QDLF$ result in the scheme being oversensitive to changes in queue behavior. The parameter $T0$ determines when the scheme starts allocating capacity to drain queues, and subsequently affects how quickly such drain capacity is allocated. In general, a large value for $T0$ is undesirable both for LANs and WANs. On the other hand, a value of $T0$ very close to zero results in increased sensitivity to changes in queues and oscillations in rate allocations. A value of $T0$ in the range $[0.5ms, 2ms]$ allows target queuing delays to be in the order of the end-to-end propogation delays for LANs, and provides quick allocation of drain capacity to clear out transient queues for WANs. The queue control parameter settings used in our simulations are $a = 1.15$, $b = 1$, $T0 = 1.5$ ms, $QDLF = 0.5$.

Since our simulations involve WAN links, we use the long-term averaging of N, load, capacity and load factor metrics. The parameters for the *long-term metric averaging* are chosen as follows. The $DecayFactor$ parameter is set to 0.9 to ensure that activity levels do not decay too quickly. The $\alpha$ parameter for long-term averaging of capacity and load factor is influenced by the fact that recent measurements have more weightage than old measurements (to allow quick response to recent events). In our simulations, $\alpha$ is set to 0.8. CCR values are copied from FRM cells, so the CCR measurement option is not used in our simulations.

In summary, the parameter set used in our simulations is:

| Parameter | Value | Purpose |
|---|---|---|
| $\delta$ | 0.1 | max-min fairness |
| $T0$ | 1.5 ms | queue control |
| $a$ | 1.15 | queue control |
| $b$ | 1 | queue control |
| $QDLT$ | 0.5 | queue control |
| $DecayFactor$ | 0.9 | long-term averaging of number of active VCs |
| $\alpha$ | 0.8 | long-term averaging of load, capacity, load factor |
| Averaging Interval | 5 ms | measurement of metrics |

## 5.2 Testing Max-Min Fairness

The first experiment we perform uses the Generic Fairness Configuration-2 (GFC-2), a popular configuration used to test the utilization, queue lengths and fairness of feedback schemes. The configuration (see figure 5) has multiple bottlenecks and connections with different round-trip times. The simulation results are shown in figure 6.

The following are the expected rate allocations as per the max-min fairness criterion. Note that the link bandwidth is adjusted by 48/53 to get an expected application throughput.

| | | | |
|---|---|---|---|
| A VCs each get 1/4 of 40 Mbps | $= 10 \times 48/53$ | $= 9.1$ | Mbps |
| B VCs each get 1/10 of 50 Mbps | $= 5 \times 48/53$ | $= 4.5$ | Mbps |
| C VCs each get 1/3 of 105 Mbps | $= 35 \times 48/53$ | $= 31.7$ | Mbps |
| D VC gets 35 Mbps | $= 35 \times 48/53$ | $= 31.7$ | Mbps |
| E VCs each get 1/2 of 70 Mbps | $= 35 \times 48/53$ | $= 31.7$ | Mbps |
| F VC gets 10 Mbps | $= 10 \times 48/53$ | $= 9.1$ | Mbps |
| G VCs each get 1/10 of 50 Mbps | $= 5 \times 48/53$ | $= 4.5$ | Mbps |
| H VCs each get 1/2 of 105 Mbps | $= 52.5 \times 48/53$ | $= 47.6$ | Mbps |

Observe that the optimal allocations are achieved in under 400 ms (under 4 round trips), and the queues are drained out within 800 ms (under 7 round trips). During the transient period, the link utilizations are close to 100% and the queue lengths are controlled to low values (maximum queue is $< 30000$ cells i.e. $< 270$ ms or 2 round trip times at 50 Mbps bottleneck rate). The steady state utilizations are close to 100% and the queue lengths are kept close to zero. The minimal oscillations in the steady state are due to the small variation in queuing delays. The initial rate assignment of each source for this simulation was picked randomly. For reasonable confidence, we repeated this experiment with other random values which gave similar results.

## 5.3 Testing Robustness: TCP Traffic and VBR Background

For testing the robustness of the scheme, we need a configuration which attacks the weaknesses of the scheme. Specifically, ERICA heavily depends upon measurement. Variation in load and capacity could lead to measurement and feedback errors resulting in unbounded queues or low average utilization. The TCP and VBR configuration (see figure 7) is designed to test this case.

The configuration simulates capacity variation by using a higher priority VBR virtual circuit which carries a multiplex of traffic from fifteen long-range dependent sources [9]. The traffic generated by this VC (and as a result, the ABR capacity) is highly variable as shown in figure 8(a). The configuration simulates load variation by using TCP sources carrying infinite ftp traffic. The load variation is caused by the startup dynamics of TCP. The TCP slow start protocol begins with small window sizes, and the amount of data it sends is limited by the window size *(window-limited)* rather than a network-assigned rate. As a result, the load offered by an individual TCP connection is bursty, i.e., it consists of active and idle periods. As the TCP window size grows, the active periods become longer. Assuming no packet losses, eventually the TCP source appears to be the same as a persistent source and its load is controlled by network-assigned rates *(rate-limited)*. The queues build up during the phase when both demand variation as well as capacity variation exist in the system. We use 100 sources and synchronize them so that the load phases (idle and active periods) of multiple sources are concentrated together. The simulation results are shown in figure 8.

Figures 8(b), (c), and (d) show ATM level metrics (ACR, queue length, and utilization) while figures 8(e) and (f) show the TCP level metrics (congestion window and sender sequence number) for three sample sources. The graphs show that ERICA successfully controls the TCP sources once they become rate-limited. As a result, the buffer requirement at the bottleneck is not a linear function of the number of sources. Though the system does not have a steady state (VBR traffic is always variable), ERICA controls the maximum and average queues and keeps utilization high (consistent with the priorities assigned in section 3.2 for graceful degradation). The congestion window and sender sequence number graphs show that the allocations to contending sources were also fair despite the variation in load and capacity.

# 6 Conclusions

In this paper, we have described the design and evaluation of the ERICA switch algorithm for ATM ABR congestion control. As a basis for the design process, we presented a simple switch model and formulated design goals. The key design goals are max-min fair steady state operation with controlled queueing delays, stability, and robustness to variation in ABR load and capacity. We then presented the ERICA algorithm, showing how the goals and simplicity determine every step in the algorithm.

The scheme entails that the switches periodically monitor their load on each link and determine a load factor, the available capacity, the queue length, and the number of currently active virtual channels. This information is used to calculate a fair and efficient allocation of the available bandwidth to all contending sources. The measurement aspects which determine the robustness of the algorithm are treated in depth. An analytical proof of convergence to steady state max-min fairness is given in appendix A. In addition, we present simulation results illustrating how the scheme meets the desired goals such as good steady state performance (high utilization, controlled queuing delay, max-min fairness), quick convergence from network transients, and robustness to load/capacity variation.

The ERICA scheme has considerably influenced the design of contemporary switch schemes. Notably, the ATM Forum traffic management specification 4.0 [1] cites ERICA as an example switch mechanism. A patent on the scheme features is currently pending [10].

# References

[1] "The ATM Forum Traffic Management Specification Version 4.0," ATM Forum Traffic Management AF-TM-0056.000, April 1996. Available as ftp://ftp.atmforum.com/pub/approved-specs/af-tm-0056.000.ps

[2] L. Kalampoukas, A. Varma, K. K. Ramakrishnan, "An efficient rate allocation algorithm for ATM networks providing max-min fairness," *In Proceedings of the 6th IFIP International Conference on High Performance Networking*, September 1995.

[3] K. Siu and T. Tzeng, "Intelligent congestion control for ABR service in ATM networks," *Computer Communication Review*, Volume 24, No. 5, pp. 81-106, October 1995.

[4] Y. Afek, Y. Mansour, and Z. Ostfeld, "Phantom: A simple and effective flow control scheme," *In Proceedings of the ACM SIGCOMM*, August 1996.

[5] D. Tsang and W. Wong, "A new rate-based switch algorithm for ABR traffic to achieve max-min fairness with analytical approximation and delay adjustment," *In Proceedings of the 15th IEEE INFOCOMM*, pp. 1174-1181, March 1996.

[6] A. Charny, D. D. Clark, R. Jain, "Congestion Control with Explicit Rate Indication," In Proceedings of ICC'95, June 1995.

[7] A. Arulambalam, X. Chen, and N. Ansari, "Allocating Fair Rates for Available Bit Rate Service in ATM Networks," *IEEE Communications Magazine*, 34(11):92–100, November 1996.

[8] R. Jain, S. Kalyanaraman and R. Viswanathan, "The OSU Scheme for Congestion Avoidance in ATM Networks: Lessons Learnt and Extensions,"[2] Performance Evaluation Journal, Vol. 31/1-2, December, 1997.

[9] S. Kalyanaraman, "Traffic Management for the Available Bit Rate (ABR) Service in Asynchronous Transfer Mode (ATM) networks" *Ph.D. Dissertation*, Dept. of Computer and Information Sciences, The Ohio State University, August 1997.

[10] R. Jain, S. Kalyanaraman, R. Goyal, R. Viswanathan, and S. Fahmy, "ERICA: Explicit Rate Indication for Congestion Avoidance in ATM Networks," *U.S. Patent Application* (S/N 08/683,871), filed July 19, 1996.

[11] A. Arora and M. Gouda, "Closure and convergence: A foundation of fault-tolerant computing" *IEEE Transactions on Software Engineering,* Vol. 19, No. 10, pages 1015 – 1027, 1993.

[12] R. Jain, "Congestion Control and Traffic Management in ATM Networks: Recent Advances and a Survey," *Computer Networks and ISDN Systems*, October 1996.

[13] R. Jain, S. Kalyanaraman, S. Fahmy, R. Goyal, "Source Behavior for ATM ABR Traffic Management: An Explanation," *IEEE Communications Magazine*, November 1996.

[14] S. Kalyanaraman, R. Jain, S. Fahmy, R. Goyal and J. Jiang, "Performance of TCP over ABR on ATM backbone and with various VBR traffic patterns," *In Proceedings of ICC'97*, June 1997.

[2] All our papers and ATM Forum contributions are available through www.cis.ohio-state.edu/~ jain

[15] R. Goyal, R. Jain, S. Kalyanaraman, S. Fahmy, B. Vandalore, S. Kota, "TCP Selective Acknowledgments and UBR Drop Policies to Improve ATM-UBR Performance over Terrestrial and Satellite Networks", *In Proceedings of ICCCN'97*, September 1997,

[16] A. Romanov, S. Floyd, "Dynamics of TCP Traffic over ATM Networks," *IEEE Journal on Selected Areas of Communications*, Vol 13, No. 4, May 1996.

[17] F. M. Chiussi, Y. Xia, and V. P. Kumar. "Dynamic max rate control algorithm for available bit rate service in ATM networks," *In Proceedings of the IEEE GLOBECOM*, volume 3, pages 2108–2117, November 1996.

[18] S. Kalyanaraman, R. Jain, J. Jiang, R. Goyal, S. Fahmy and P. Samudra, "Design Considerations for the Virtual Source/Virtual Destination (VS/VD) Feature in the ABR Service of ATM Networks," *Submitted to IEEE INFOCOM 1998.*

[19] S. Fahmy, R. Jain, R. Goyal, B. Vandalore, S. Kalyanaraman, S. Kota and P. Samudra, "Feedback Consolidation Algorithms for ABR Point-to-Multipoint Connections in ATM Networks," *To appear in IEEE INFOCOM 1998.*

[20] N. Plotkin and J. Sydir, "Behavior of Multiple ABR Flow Control Algorithms Operating Concurrently within an ATM Network," *In Proceedings of IEEE INFOCOM 1997*, Kobe, April 1997.

[21] H. Özbay, S. Kalyanaraman, A. İftar, "On Rate-Based Congestion Control in High Speed Networks: Design of an $\mathcal{H}^\infty$ Based Flow Controller for a Single Bottleneck," *Submitted to American Control Conference, 1998*[3].

[22] R. Jain, "The Art of Computer Systems Performance Analysis," *John Wiley & Sons*, 1991.

# A Proof of Convergence to Max-Min Fairness

In this appendix we give an analytical proof for the convergence of a single bottleneck node implementing ERICA towards max-min fair rate allocations. We make the following assumptions about the system:

[3] Available through www.ecse.rpi.edu/Homepages/shivkuma

- Single bottleneck node
- Use of the "core" ERICA algorithm (defined in section 4.2) with two exceptions:
  1. We ignore the effect of the queue control function.
  2. We ignore the moderation step (equation (11)):
     IF (ER > Fairshare AND CCR < Fairshare) THEN ER = Fairshare
- Sources are persistent (always have data to send), though some (not all) might be source-bottlenecked at low rates
- Heterogeneous round-trip times
- Switch averaging interval is at least the maximum of **a)** the largest RTT of any VC though the bottleneck, and **b)** the time required to see at least one RM cell of every active source (inter-RM cell time). Basically, we assume that measurements are reliable since they are made over sufficiently long time scales.
- A *"cycle"* is equal to a switch averaging interval as defined above.
- Load factor ($z$) > 0 and ER < Link Rate
- Source-bottleneck behavior (if any) does not change during the convergence period.

**Notation:**

- Rate of source $i$ in cycle $j$ (CCR) is $R(i, j)$
- MaxAllocPrevious in cycle $j$ is $Max_i\ R(i, j)$
- The ER for source $i$ in cycle $j$ is the same as the rate of source $i$ in cycle $j + 1$, i.e., $R(i, j + 1)$
- $z_j$ = overload factor measured in $j$th cycle (and used in $(j + 1)$ th cycle).
- $C$: Target ABR capacity of the bottleneck.
- $B$: Sum of the rates of bottlenecked sources, also equal to $b \times C$, $b \leq 1$

- $N$: Number of active sources

**Definition:** A source is said to be *satisfied* at a given rate if it is bottlenecked elsewhere and cannot utilize higher rate allocations.

**To prove:** that for the system described above, the ERICA algorithm causes it to converge towards max min operation in at most O(logN) number of cycles.

**Proof:**

The proof methodology used here was proposed in reference [11]. We first prove a set of *safety (closure)* properties which show that the system remains within a closed state space, S. Then we prove a set of *convergence properties* which show that the system reaches and remains in a target state space, T.

We split the state space based upon the initial value of the overload factor, z i.e., **a)** z < 1 **b)** $z \geq 1$

The closed state space, S is:
S: $0 < z < N$
The target convergence state space, T is:
T: $(1 \leq z \leq 1 + \delta)$ AND Allocations are Max-Min fair,

where the term "Max Min fair" implies that contending sources are allocated the highest possible *equal* rates, satisfying the condition on $z$.

**Closure Properties:**

**Lemma 0:** Given that the maximum rate (C) of any VC is at most the target link rate, the overload factor lies between 0 and N, where N is the number of VCs set up (assumed active).

This is proved trivially given the system assumptions enumerated earlier. □

**Convergence Properties:**

**Lemma 1:** ERICA takes one cycle to satisfy sources bottlenecked at rates below equal Fairshare (C/N).

**Proof:** In every cycle, ERICA allocates at least Fairshare = "fs" = C/N to every source. If there exist sources which are bottlenecked such that they cannot utilize rate allocations above fs, the system *satisfies* such sources in one cycle. This first cycle is called *"initialization cycle"* in what follows. □

**Note 1:** During convergence, there is at most one initialization cycle for any configuration.

**Note 2:** After the VCs below fs are satisfied, the unused capacity (if any) will be reflected in the value of the overload factor, $z$ (which is the ratio of the total load and the target capacity).

**Note 3:** The following lemmas assume that the initialization cycle is completed, and that there is at least one "greedy" or "unconstrained" source going through each bottleneck which can utilize any bandwidth allocated to it.

**Lemma 2:** If a switch is underloaded, i.e., $z < 1$, then in O(log(N)) cycles, either the system converges to the target state space, T, or the load factor increases to reach a value greater than unity.

**Proof:**

During underload ($z < 1$), ERICA uses the following formula to allocate rates:
ER = Max(MaxAllocPrevious, $CCR/z$).

Recall that ER = $R(i,j)$, MaxAllocPrevious = $Max_i\ R(i,j-1)$, and $CCR/z = R(i,j-1)/z_{j-1}$

Hence, the ERICA formula can be rephrased as:

$$R(i,j) = Max(\ Max_i\ R(i,j-1),\ R(i,j-1)/z_{j-1}\ ) \tag{13}$$

Note that MaxAllocPrevious ( $Max_i\ R(i,j-1)$ ) is at least $C/N$ (equal to the maximum of the allocations in the previous cycle) and $CCR/z$ is greater than $CCR$. As a result, the allocation of *every unsatisfied source increases.*

If all sources are greedy and and initially equal, the new load factor is unity, with all sources equal. *In this case the target $T$ is achieved in a single cycle.*

In the case that source rate allocations are unequal and/or some sources are satisfied, the behavior of the system is different. Satisfied sources stay constant and the overload factor increases in the next cycle. If all sources are greedy, they get a rate of C/N in the first cycle. As a result, the new load factor is at least load/capacity = $(N \times (C/N))/C = 1$. *In this case, the load factor becomes greater than unity in a single cycle.*

We now show that even if the above special conditions do not hold, the load factor becomes greater than unity in O(log N) cycles. Assume that some sources at bottlenecked at rates below $C/N$, and the sum of their rates is B. The remaining sources get at least the maximum allocation of the previous cycle, i.e. $Max_i\ R(i,j-1)$. Starting from an initial load factor of $z_0$, the system increases its load factor in every cycle. Assume that, in the $(j-1)$ th cycle the overload factor, $z_{j-1}$ is less than $1/(1+\epsilon)$, for small $\epsilon$. Now,

$$z_j = \frac{B + \sum_i R(i,j)}{C} \geq \frac{B + \sum_i \frac{R(i,j-1)}{z_{j-1}}}{C} \qquad \text{from (13)}$$

$$\geq \frac{B + \sum_i \frac{R(i,j-2)}{z_{j-2} \times z_{j-1}}}{C} \geq \frac{B + \sum_i \frac{R(i,0)}{z_0 \times z_1 \times \ldots \times z_{j-1}}}{C} \geq \frac{B + \sum_i \frac{R(i,0)}{(1/1+\epsilon)^j}}{C}$$

For $z_j$ to become greater than 1, it is sufficient that:

$$B + \sum_i \frac{R(i,0)}{(1/1+\epsilon)^j} > C, \quad \text{i.e.,}$$

$$j < \left| \log_{1+\epsilon} \sum_i \frac{R(i,0)}{C-B} \right|$$

Since $B$ and $R(i,0)$ are constants, and C is upper bounded by the link capacity $j = O(logN)$ in the worst case.

**Note 1:** $z_j$ can also become greater than 1 when:

$B + \sum_i Max_i \; R(i,j-1) = B + (N-Nb) \times Max_i \; R(i,j-1) > C$, where Nb is the number of bottlenecked sources. Here, we have taken the $Max_i \; R(i,j-1)$ term in the ERICA step given in equation (13) instead of the $R(i,j-1)/z_{j-1}$ term which is used in the above proof. This new inequality reduces to:
$Max_i \; R(i,j-1) > (C-B)/(N-Nb)$

Observe that the right hand side of the above inequality is the target max-min rate allocation, which means that $z_j$ becomes greater than unity in *one cycle* when *any one* of the rates $R(i,j-1)$ is greater than the final max-min allocation. Note that this assumes that the moderation step (see list of assumptions) has been ignored. □

**Lemma 3:** If a switch is overloaded, i.e., $z \geq 1$, then the switch remains overloaded i.e. $z \geq 1$, and converges within O(logN) cycles to the desired operating region, T.

**Proof:** We split the proof into three parts:

**Part A:** We first prove that the system remains in the region $z \geq 1$.

With the system starting at $z_{j-1} \geq 1$, we show that the minimum value of the new load factor after a cycle, $z_j$, is greater than or equal to unity.

The ERICA code segment used for this proof is:

IF $(z \leq 1+\delta) ER = Max$(MaxAllocPrevious,$CCR/z$)
    ELSE $ER = Max(C/N, CCR/z)$

We argue that the ER value obtained by the assignment statement $ER = Max$(MaxAllocPrevious,$CCR/z$) does not reduce the load factor below its current value. Recall that MaxAllocPrevious $= Max_i\ R(i, j-1)$ and $CCR/z = R(i, j-1)/z$. Now, since $z \geq 1$, MaxAllocPrevious $\geq CCR/z$. As a result, this term is not going to reduce $z$. Therefore, we simply deal with the second assignment statement in the ERICA code segment above, i.e., $ER = Max(C/N, CCR/z)$.

Split the set of sources into two categories:

1. Sources bottlenecked at rates equal to or below $C/N$, which have a total rate of $b \times C, b \geq 0$.

2. Sources above $C/N$, with a total rate of $d \times C$.

The current load factor is $z_{j-1} = ((b+d) \times C)/C = b + d > 1$. If all sources were to divide their rates by $z_{j-1}$, the new load factor $z_j$ would be unity. In our case only sources above $C/N$ reduce their rates. The new load factor is $b + (d/z_{j-1})$. To complete the proof of part A, note that:

$z_j = b + (d/z_{j-1}) \geq (b+d)/z_{j-1} = 1$ □

**Part B:** In the worst case, the system first reaches the region $1 \leq z < 1+\delta$ in O(logN) cycles.

If the system is already in region $1 \leq z < 1+\delta$, the proof is trivial.

Else, let the initial load factor be $z_0$ and the current load factor be $z_j$. Let $B = b \times C$ be the sum of bottlenecked rates at or below C/N. The remaining rates $R(i,j) \geq C/N$, and $z_k > 1+\delta$, $\forall k < j$. A technique similar to the one shown in lemma 2 can be used to prove that $j = O(logN)$, i.e., the system reaches the operating region $1 \leq z < 1+\delta$ in O(log N) cycles. □

**Part C:** The contending sources get an equal rate allocation in the region $1 \leq z < 1+\delta$

The ERICA allocation in this region (in the $j+1$ th cycle) is: Max(MaxAllocPrevious, $CCR/z$) i.e. $Max(\ Max_i\ R(i,j),\ R(i,j)/z_j\ )$

Since $z_j \geq 1$, $R(i,j)/z_j < Max_i\ R(i,j)$, and the ERICA allocation is simply $Max_i\ R(i,j)$ for all sources. In other words, the rate allocations to all sources in this region are equal.

**Note 1:** Observe that if $R(i,j)$ s were already equal, the load factor would be unchanged in subsequent cycles, i.e., the system would remain at $1 \leq z < 1+\delta$, and rates of contending sources $R(i,j)$ are equalized, leading to max-min fair allocations. That is, the system has reached the state, T, and stays in this state until new input changes occur.

If the rates $R(i,j)$ are not equal before this *"equalization cycle"*, the new load factor can be greater than $1+\delta$. As proved in part B, the system requires at most O(log N) cycles to converge to the state where $1 \leq z < 1+\delta$. However note that at every cycle of this aforementioned convergence process, all rate allocations remain equal since they are scaled by the same factor ($z$). This implies that the system has reached a state where $1 \leq z < 1+\delta$ AND all rate allocations of unconstrained sources are equal. But this state is the same as the target state space, T.

**Theorem 1:** From an arbitrary initial state, the ERICA algorithm brings the system to the target operating region T within O(log N) cycles.

An arbitrary initial state can be characterized by a value of the load factor $z$ between 0 and N (closure, lemma 0). If $z < 1$, we have shown in lemma 2 that the system reaches a state where $z \geq 1$ within O(log N) cycles. Once $z \geq 1$, we have shown that the load factor does not reduce below unity (lemma 3, part A). Further, the system moves to the region $1 \leq z < 1+\delta$ within O(log N) cycles (lemma 3, part B) and the rates are equalized in a single in this region (lemma 3, part C). The system may now remain stable in the region $1 \leq z < 1+\delta$, with equal rates (i.e. max-min fair allocations), or move out of the region and converge back and remain in this region in O(log N) cycles with the rates being equal at every cycle during this convergence process (lemma 3, part C, note 1). This final region of stability is in fact the target state space, T, i.e., $1 \leq z < 1+\delta$, and allocations are max-min fair. The maximum number of cycles to converge to T from an arbitrary initial state is O(log N). □

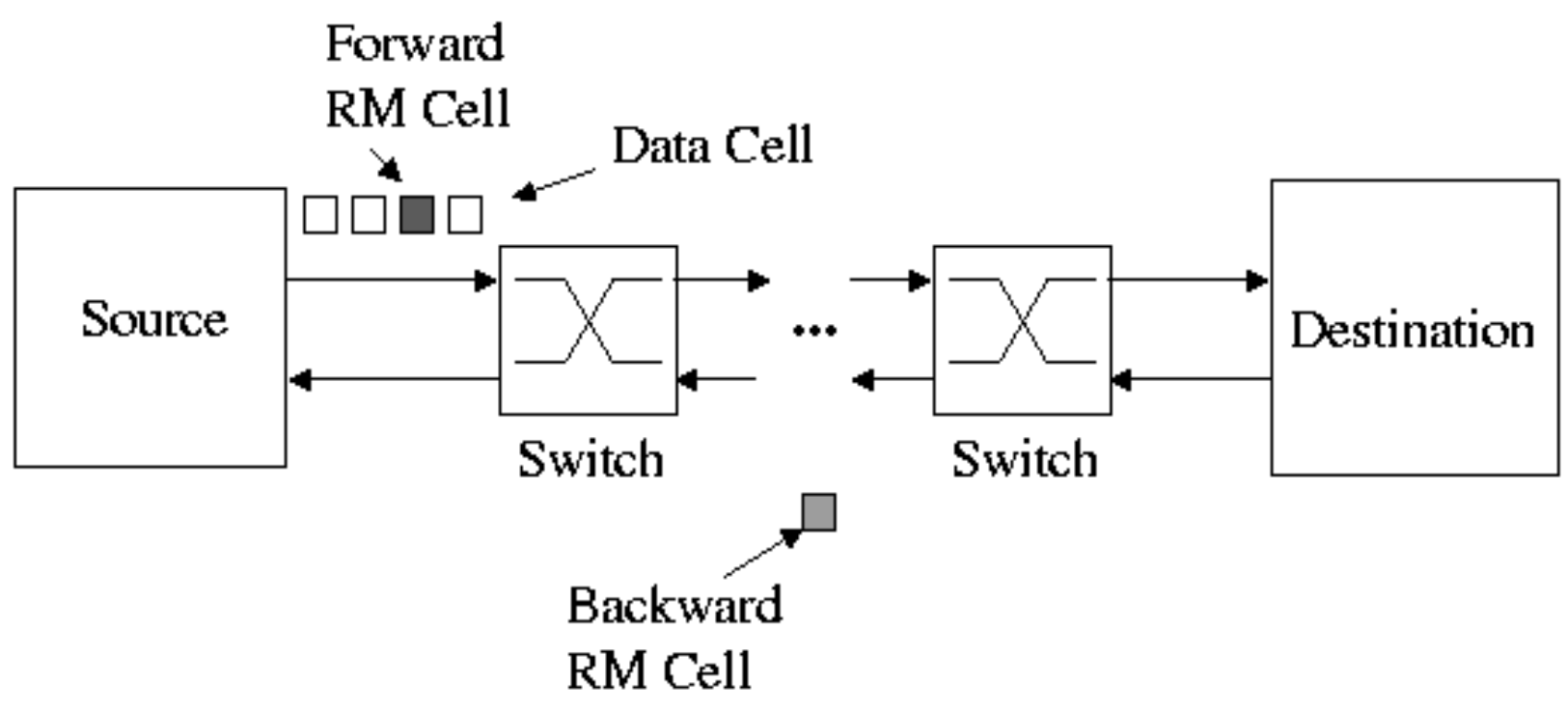

Figure 1: RM cell path

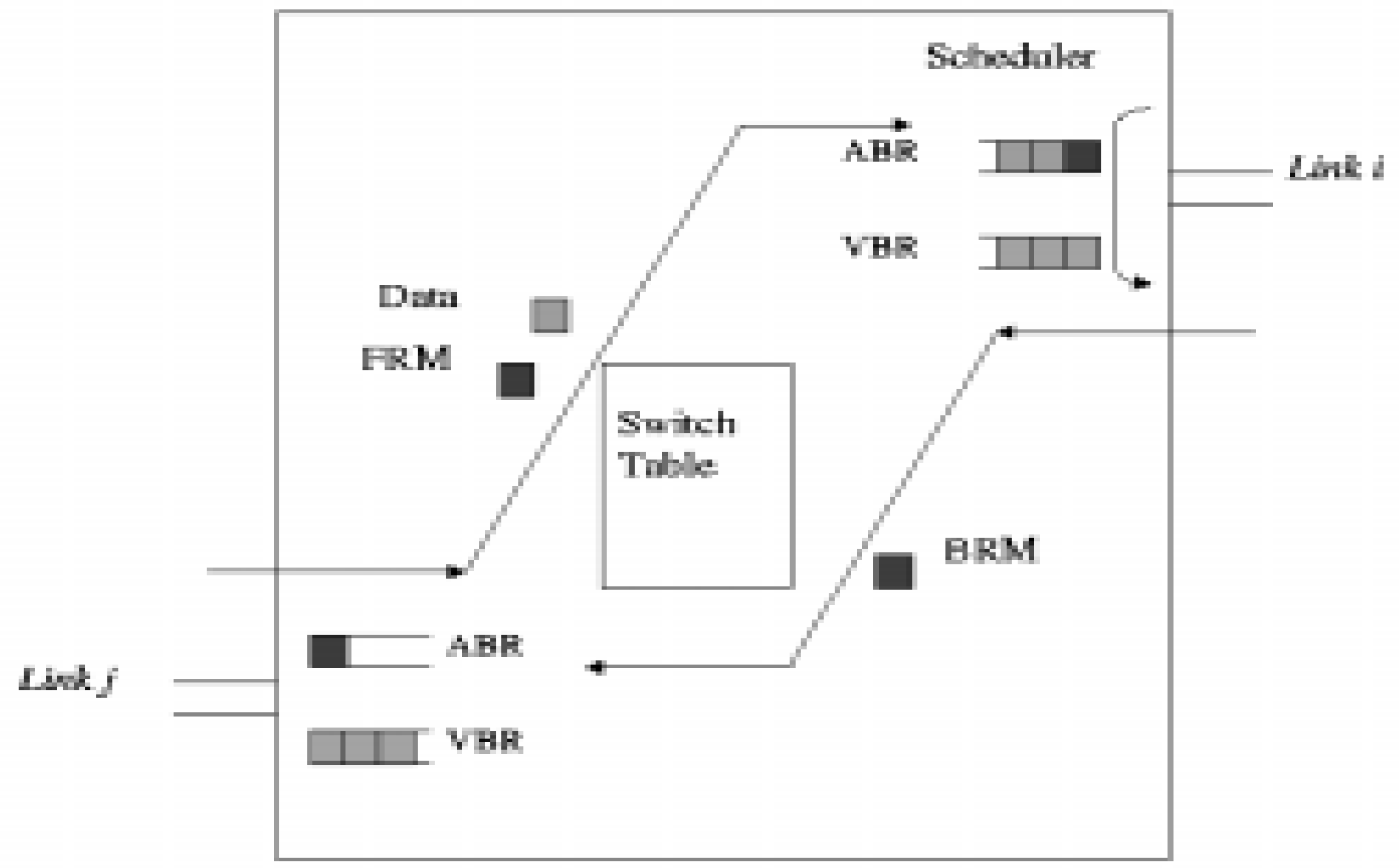

Figure 2: Switch model

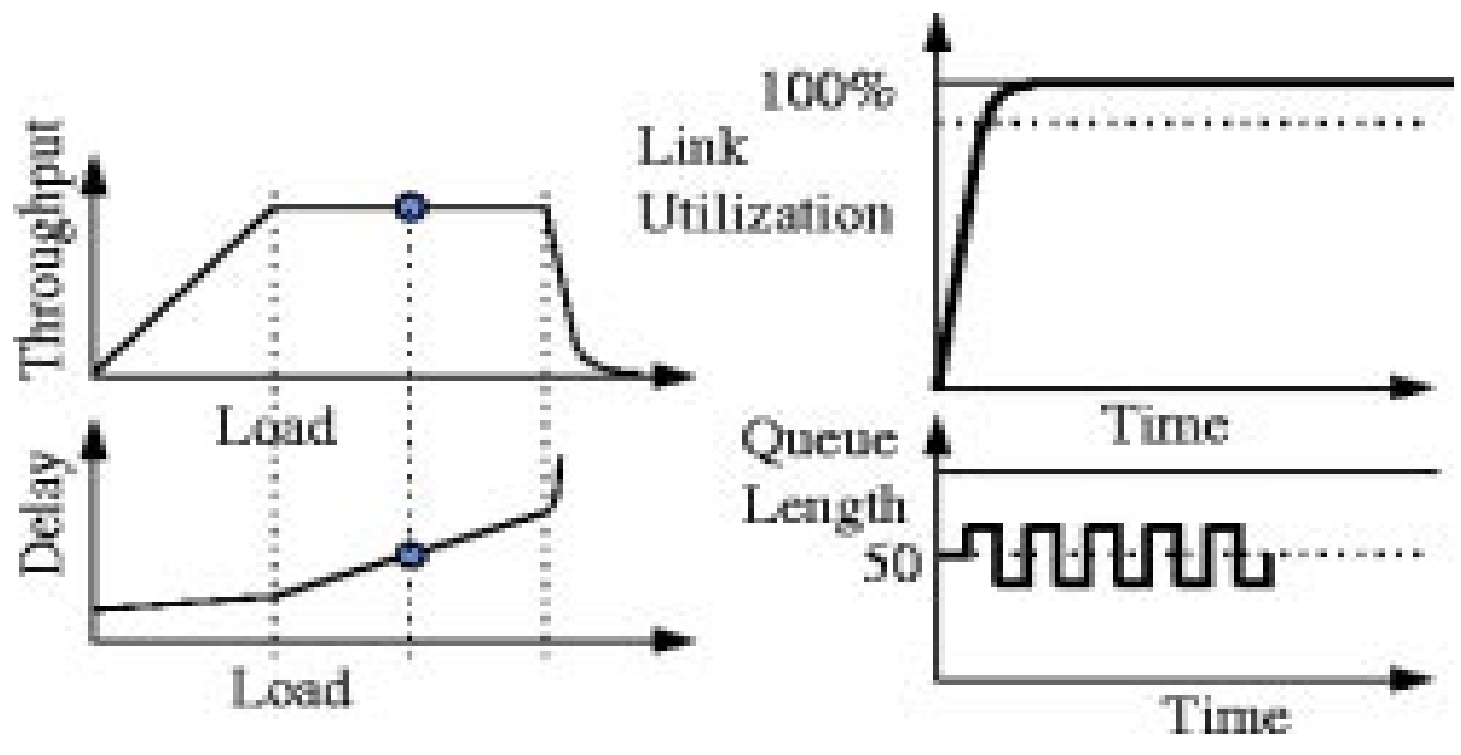

Figure 3: Target throughput-delay behavior of the system

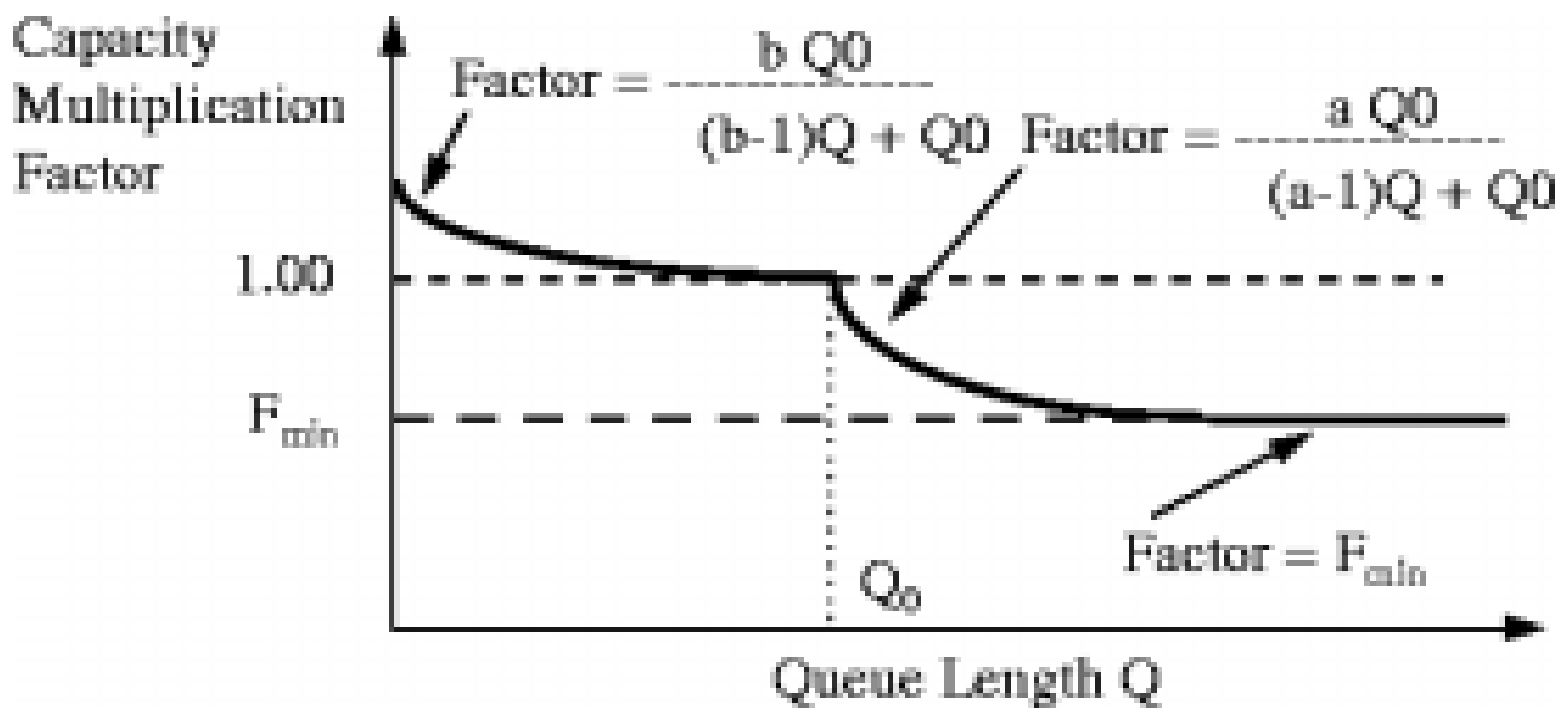

Figure 4: The queue control function in ERICA

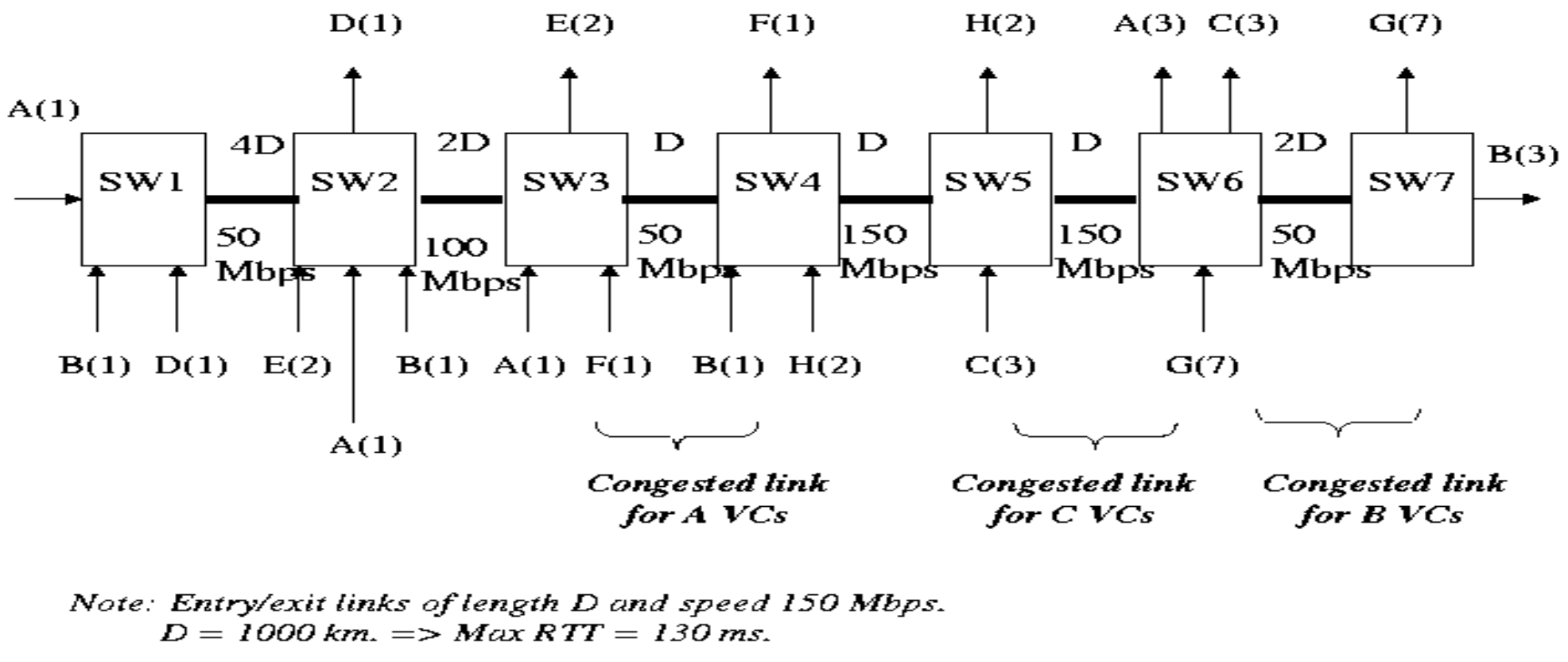

Figure 5: The Generic Fairness Configuration-2

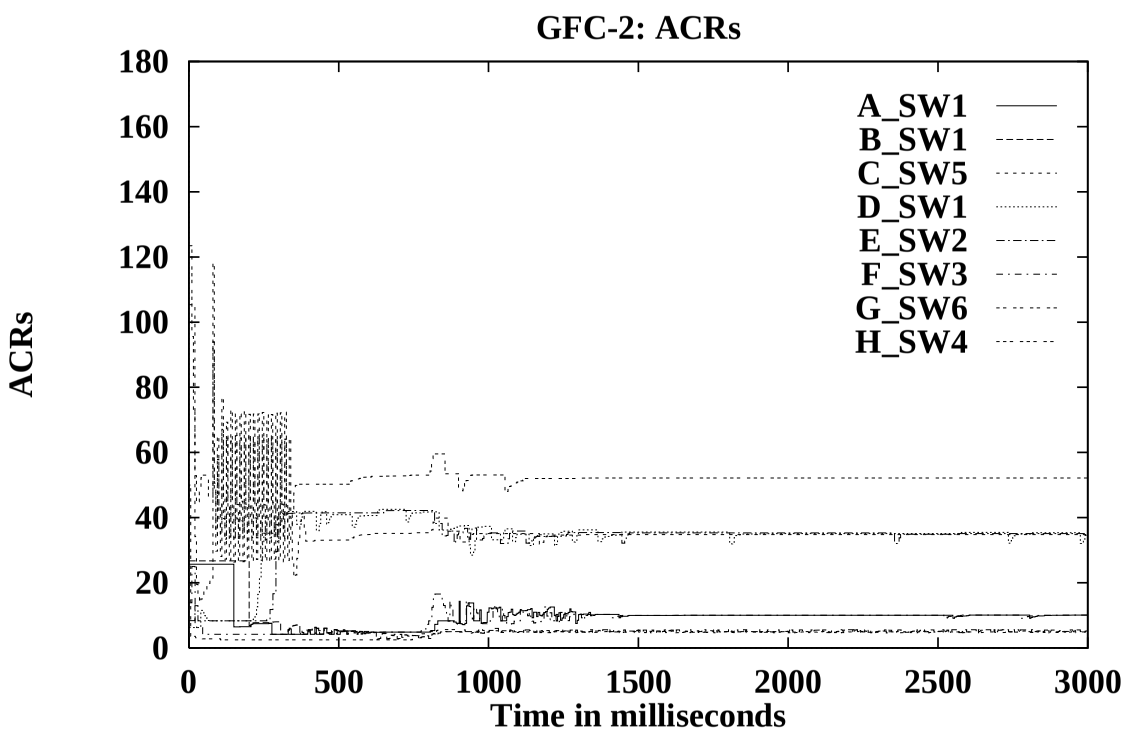

(a) Allowed Cell Rate (ACR)

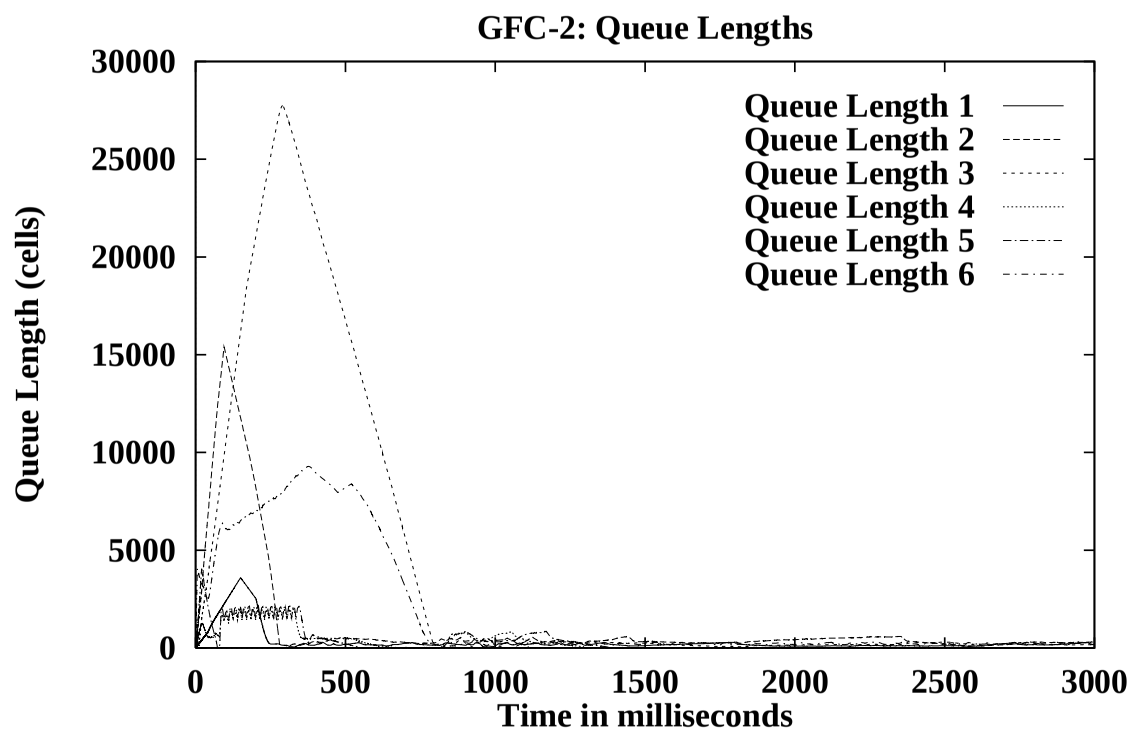

(b) Bottleneck Queue Lengths

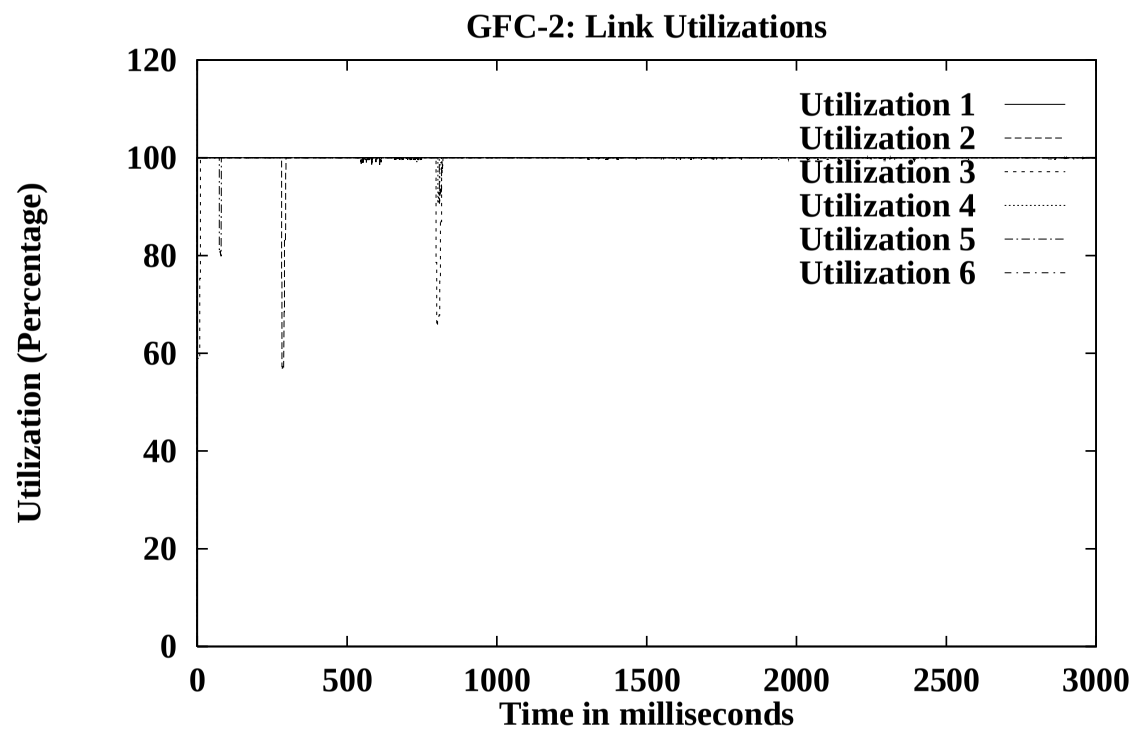

(c) Bottleneck Link Utilizations

Figure 6: Simulation results with the GFC-2 configuration

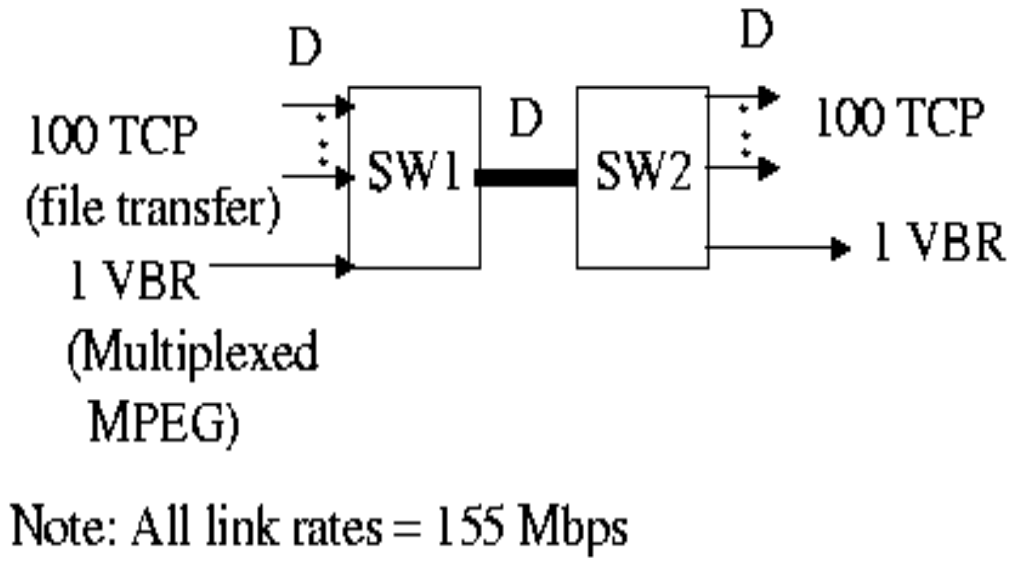

Figure 7: TCP and VBR configuration

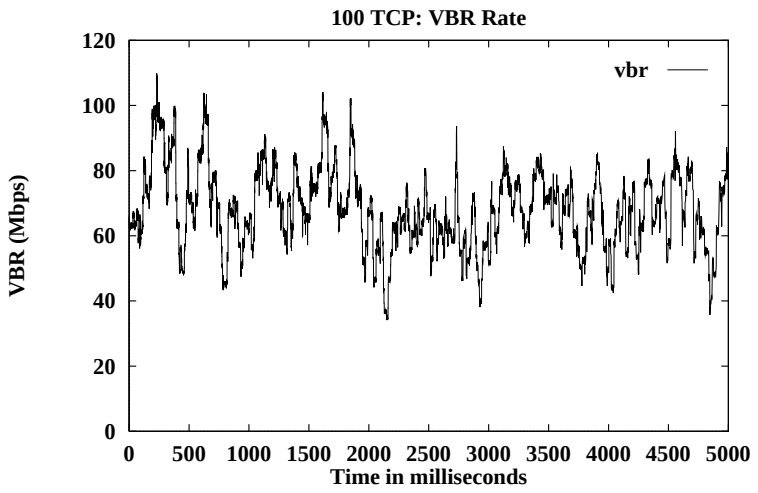

(a) VBR Rate

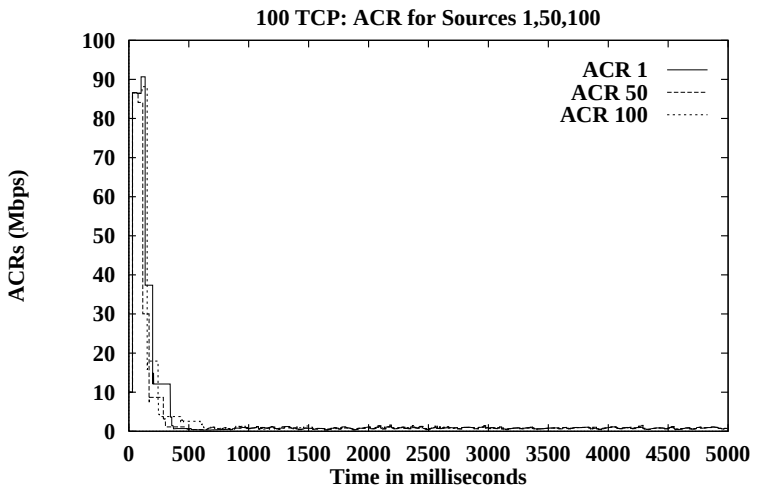

(b) Allowed Cell Rate (ACR)

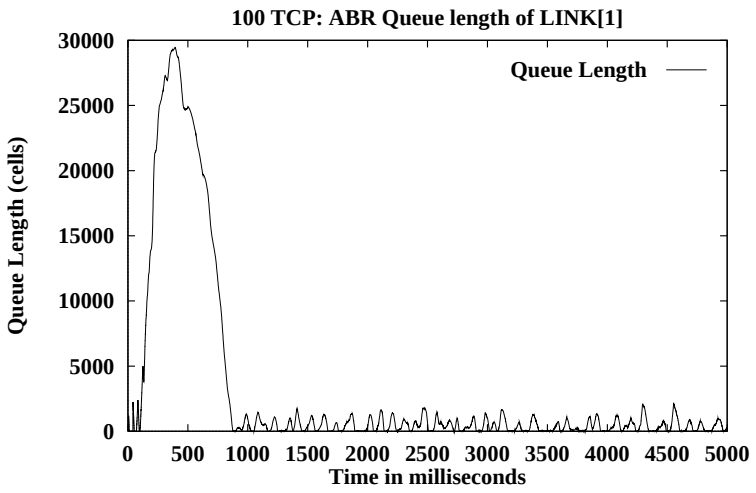

(c) Bottleneck Queue Length

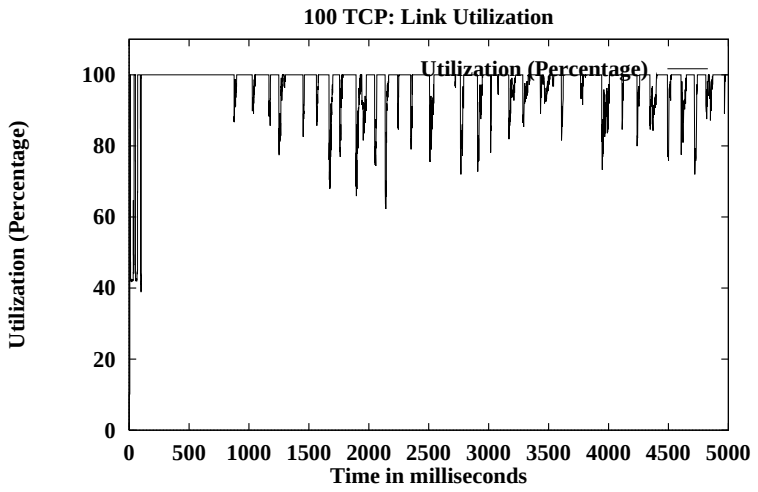

(d) Bottleneck Link Utilization

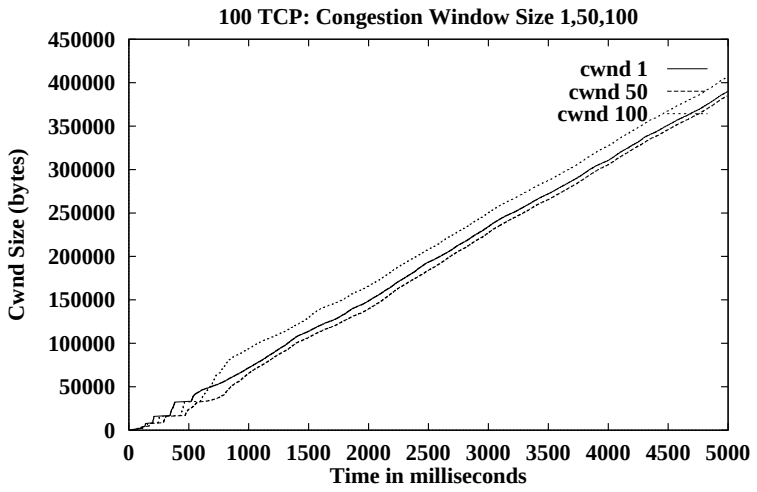

(e) TCP Congestion Window

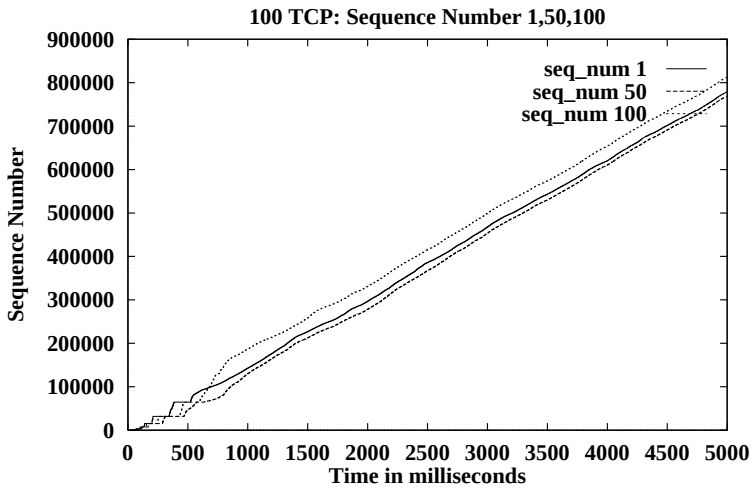

(f) TCP Send Sequence Numbers

Figure 8: Simulation results with the 100 TCP and VBR configuration